\documentstyle[preprint,aps]{revtex}
\begin{document}
\draft
\thispagestyle{empty}
\preprint{IP--BBSR--95/15 }
\title{\bf STRINGS IN A WORMHOLE BACKGROUND }
\author{Sayan Kar \thanks{ Electronic Address
: sayan@iopb.ernet.in}}
\address{Institute of Physics, Sachivalaya Marg \\
Bhubaneswar, 751005 INDIA}
\maketitle
\begin{abstract}

Exact solutions of the string equations of motion in a specific
Lorentzian wormhole
background are obtained. These include both closed and open
string configurations. Perturbations about some of these
configurations are investigated using the manifestly covariant
formalism of Larsen and Frolov. Finally,
the generalized  Raychaudhuri equations for the corresponding
string worldsheet deformations are written down and analysed
briefly.

\end{abstract}

\pagebreak

\section{\bf INTRODUCTION}

The study of the string equations of motion and constraints in
a curved background spacetime has been a topic of active
research over the last few years. Since the equations are
nonlinear it is often quite difficult to obtain exact solutions
. However, presently, there exists several exact solutions in
a variety of curved backgrounds. These include solutions in
the spacetime around a cosmic string [1], De Sitter space [2], black
hole geometries [3], cosmological backgrounds [4], and
gravitational wave backgrounds [5]. More interestingly, several
multi--string solutions have been found in De Sitter space [6].
 The aim of this paper is to
analyse the string equations of motion and constraints in the
background of a Lorentzian wormhole geometry, obtain some
exact string configurations and study their properties related
to perturbations.

If the background spacetime is static and we consider only
stationary strings there is a nice simplification of the
equations of motion. It turns out that one essentially
has to solve for the geodesics in a certain {\em unphysical}
three dimensional Riemannian space in order to obtain specific string
configurations. This fact was first put to use by Frolov et. al. [3]
to obtain string configurations in a Kerr--Newman background
and we shall also exploit it in our investigations here.
Our analysis, as we shall see, will reveal interesting open as
well as closed string configurations. The case of the closed
string is particularly important because in black
hole backgrounds such solutions are {\em not} possible [7].
Furthermore, following a recent analysis due to Larsen and
Frolov [8] we shall investigate the perturbations about these
string configurations. Lastly, we shall write down the
generalized Raychaudhuri equations [9] for worldsheet deformations
and comment briefly on its solutions and the issue of worldsheet
focussing.

Since a wormhole geometry will be used throughout as the
background spacetime it is probably necessary to say a few words
on wormholes in general. Lorentzian signature wormholes
have been a topic of great interest in recent times primarily
because of two reasons. Firstly, these geometries require
the violation of the Energy conditions for the matter that
is required to support them [10]. One does not know till today
whether such matter is possible although quantum stress energy
seems to be natural choice. Secondly, such wormholes can very
easily be converted into a time machine by performing a relative
motion of the wormhole mouths [11]. This makes the possibility of going back
to ones past a seemingly simple notion and a large number
of physicists have spent a lot of time in trying to understand
the consequences of such backward time travel within the
framework of simple models [12].

The organisation of the paper is as follows. Section II contains
a discussion of the string equations of motion and the constraints
and derives the various possible solutions representing string
configurations. In Section III we derive the perturbation
equations and solve them for some cases. The Raychaudhuri
equations are written down and analysed in Section IV. Section
V is a conclusion with comments on possible future directions.
The Appendix to the paper lists the various Affine Connections and Riemann
tensors
used extensively in the calculations.

Units and sign conventions in this paper are those due to Misner,
Thorne and Wheeler [13]. ${(2\pi \alpha)}^{-1}$ , where $\alpha$
is the inverse of the string tension is set to one.

\section{\bf STRING EQUATIONS OF MOTION, CONSTRAINTS AND THEIR
SOLUTIONS}

As mentioned earlier, the background spacetime for all the
analyses in this paper will be that of a Lorentzian wormhole
. The metric for such a geometry  is represented as:

\begin{equation}
ds^{2}= - {\chi}^{2}(l)dt^{2} + dl^{2} + r^{2}(l)({d\theta}^{2} +
{\sin}^{2}{\theta}{d\phi}^{2})
\end{equation}

where $r(l)$ and $\chi (l)$ are two functions which characterize the
nature of the geometry. The metric has the features of a wormhole
if $r(l)$ and $\chi (l)$ satisfy the following requirements :

\begin{eqnarray}
(i)\quad  r(l) \sim l \quad  as \quad  l \rightarrow \pm \infty \quad
(Asymptotic \quad flatness)\\
(ii)\qquad r(l=0) = b_{0} \quad  (\quad Existence \quad  of \quad a
\quad throat)\\
(iii)\quad \chi (l) \quad  finite \quad  everywhere \quad
(Nonexistence \quad  of \quad  Horizons)
\end{eqnarray}

Spacelike sections when embedded in a higher dimensional
Euclidean space resemble two asymptotically flat regions
connected by a bridge.

 The wormhole metric can also be written in an alternative form
by using the radial coordinate $r$ instead of the proper radial
distance $l$. This is given as :

\begin{equation}
ds^{2} = -e^{2\Phi}(r)dt^{2} +\frac{dr^{2}}{1-{\frac{b(r)}{r}}}
+ r^{2}\left (d{\theta}^{2} + {\sin}^{2}\theta d{\phi}^{2}\right
)
\end{equation}

where $\frac{b(r)}{r}\le 1$ , $b(r=b_{0}) = b_{0}$ ,
$\frac{b(r)}{r} \rightarrow 0 $ as $r\rightarrow \infty$
and $e^{2\Phi(r)}$ is always finite.

For the special case
of an Ellis geometry [14] we have $\Phi(r) = 1$ and $b(r) =
\frac{b_{0}^{2}}{r}$
or $r(l) =\sqrt {b_{0}^{2} + l^{2}}$ and $\chi (l)=1$.

Some of the interesting features of the Ellis geometry are:

(i) The matter required to support it violates all versions of
the Energy Conditions [15]. Infact $\rho (l)$ i.e. the energy
density is negative everywhere.

(ii) Exact solutions of the massless scalar wave equation can be
obtained for both the $2+1$ and $3+1$ dimensional versions of
the metric. These involve the Modified Mathieu and the Radial
Oblate Spheroidal Functions in $2+1$ and $3+1$ dimensions
respectively [16].

(iii) the $t=$ constant $\theta = \frac{\pi}{2}$ sections when
embedded in $R^{3}$ represent a catenoid which is a minimal
surface ( $Tr(K_{ij})=0$).

We now move on to the analysis of string configurations in the
background of the Ellis geometry.

The bosonic string equations of motion and constraints in a general
curved background are obtained by extremizing the Nambu--Goto
action. These are given as :

\begin{equation}
{\ddot {x^{\mu}}} - {x^{{\mu}{\prime\prime}}} +
{\Gamma}^{\mu}_{{\rho}{\sigma}} \left ( {\dot {x^{\rho}}}
{\dot {x^{\sigma}}} - {x^{{\rho}{\prime}}}{x^{{\sigma}{\prime}}}
\right ) = 0
\end{equation}

\begin{equation}
{g}_{\mu \nu}{\dot {x^{\mu}}}{x^{\nu \prime}} =
 {g}_{\mu \nu}{\dot {x^{\mu}}}{\dot x^{\nu}}
+{g}_{\mu \nu}{ {x^{\mu \prime}}}{x^{\nu \prime}}
 = 0
\end{equation}

where the primes denote derivatives with respect to $\sigma$ and
dots denote derivatives with respect to $\tau$ ($\sigma$, $\tau$
being the worldsheet coordinates).

For the case of a stationary string in a static background we
assume:

\begin{equation}
t = \tau \qquad x^{i} = x^{i}(\sigma)
\end{equation}

The string equations and constraints reduce to the following
equations :

\begin{equation}
{x^{{i}{\prime\prime}}} +
{{\Gamma}}^{i}_{{j}{k}}
 {x^{{j}{\prime}}}{x^{{k}{\prime}}}
- {\Gamma}^{i}_{00} = 0
\end{equation}

\begin{equation}
g_{00} + g_{ij}{x^{i\prime}}{x^{j\prime}} = 0
\end{equation}

where ${{\Gamma}}^{i}_{jk}$ are the Affine Connections
for the background metric.

One can also write (9) as a geodesic equation in a Riemannian space
endowed with a metric

\begin{equation}
d{\tilde s}^{2} = {H_{ij}}dx^{i}dx^{j}
\end{equation}

where $H_{ij}x^{\prime i}x^{\prime j} = 1$. The $H_{ij}$ are
related to the background metric by the following : $g_{00} = -\chi ^{2}$
; $g_{ij} = \frac{H_{ij}}{{\chi}^{2}}$ and $g_{0i} = 0$.

We shall mostly be concerned with the special case of  Ellis geometry
except while discussing closed strings. Note that we can
 reduce the problem of obtaining string configurations to that
of finding the geodesics in a fictitious Riemannian space. However
 we have to make sure that the constraint equations are satisfied.

For our metric, i.e Ellis geometry the corresponding components
of $H_{ij}$ are fairly simple.

\begin{eqnarray}
H_{11} = 1 \qquad H_{22}=r^{2}(l) = b_{0}^{2} + l^{2} \qquad \\
H_{33} = r^{2}(l)\sin^{2}{\theta} = (b_{0}^{2} + l^{2})\sin^{2}{\theta}
\qquad H_{ij} = 0 \forall i\neq j
\end{eqnarray}

The equations of motion now reduce to the following geodesic
equations:

\begin{eqnarray}
l^{\prime \prime} - {r\tilde r}{{\theta}^{\prime}}^{2} -
{r\tilde r}{\sin ^{2}{\theta}}{{\phi}^{\prime}}^{2} = 0 \\
{\theta}^{\prime \prime} + 2{\frac{\tilde r}{r}}{\theta
^{\prime}}{l^{\prime}} - \sin {\theta} \cos {\theta} {\phi
^{\prime}}^{2} = 0 \\
{\phi}^{\prime \prime} + 2{\frac{\tilde r}{r}}{\phi ^{\prime}}
l^{\prime} + 2\cot {\theta} {\theta ^{\prime}}{\phi ^{\prime}} =
0
\end{eqnarray}

In the above the $\sim$ sign denotes differentiation w.r.t
the variable $l$.

The constraint equation leads to the following :

\begin{equation}
-1 + {l^{\prime}}^{2} + r^{2}\left ( {{\theta}^{\prime}}^{2} + {\sin
^{2}{\theta}}{{\phi}^{\prime}}^{2} \right ) = 0
\end{equation}

We now specialize by choosing $\theta = \frac{\pi}{2}$. The
second of the geodesic equations is solved straightaway and
therefore we need to look for solutions of only the $l$ and
$\phi$ equations. The $\phi$ equation can be integrated once to
give:

\begin{equation}
{\phi}^{\prime} = \frac{C_{1}}{r^{2}}
\end{equation}
where $C_{1}$ is an integration constant. Integrating the $l$
equation once by making use of the above expression for $
{\phi}^{\prime}$ one gets :
\begin{equation}
-C_{2} + {l^{\prime}}^{2} + {\frac{C_{1}^{2}}{r^{2}}} = 0
\end{equation}

The above equation matches with the constraint equation only if
we choose $C_{2} = 1$. We can now obtain the various string
configurations by solving the two equations for $\phi ^{\prime}$
and $l^{\prime}$.

\subsection{\bf  General Solution for $\theta =
\frac{\pi}{2}$}

The solutions of the equations (18) and (19) are as follows :

\begin{equation}
\sigma - {\sigma}_{0} = b_{0}F(\beta , q) -
b_{0}E(\beta , q) + l{\sqrt{\frac{b_{0}^{2} + l^{2}}{a_{0}^{2} +
l^{2}}}}
\end{equation}

where ${\sigma}_{0}$ is an integration constant, $a_{0}^{2} =
b_{0}^{2} - l^{2} $ and
\begin{equation}
\beta = {\arctan {\frac{b_{0}}{l}}} \qquad , \qquad q =
{\frac{C_{1}}{b_{0}}}
\end{equation}

 $F(\beta , q)$ and $E(\beta , q)$  are the Elliptic
Functions of the first and second kinds given by:

\begin{equation}
F(\beta,k) = \int ^{\beta}_{0}
\frac{d\alpha}{\sqrt{1-k^{2}\sin^{2}\alpha}}
\end{equation}

\begin{equation}
E(\beta,k) = \int ^{\beta}_{0}
{d\alpha}{\sqrt{1-k^{2}\sin^{2}\alpha}}
\end{equation}

with $k^{2}$ less than one .

The solution for $\phi$ as a function of $l$ is also given in
terms of Elliptic functions.

\begin{equation}
\phi - {\phi}_{0} = {\frac{C_{1}}{b_{0}}}F(\alpha , q)
\end{equation}

where

\begin{equation}
\alpha = {\arctan {\frac{l}{a_{0}}}} \qquad , \qquad
q=\frac{C_{1}}{b_{0}}
\end{equation}

Much simpler functional forms can be obtained by considering
special cases of the above expressions. These are discussed
below.

\subsection{\bf Straight Strings}

{}From the original string equations of motion and the constraints
one can easily see that the following simple configuration
represents a solution.

\begin{equation}
l = \sigma \qquad \theta = \frac{\pi}{2} \qquad \phi =
{\phi}_{0}
\end{equation}

This represents a straight stringin the sense that the proper
radial distance $l$ is identified with the worldsheet
cooordinate $\sigma$ and $\phi$ does not vary throughout.

\subsection{\bf Strings With $\phi \neq Constant$}

A special case of the general solution described in the Case 1
can be obtained by setting $a_{0} = 0$ which implies $b_{0} =
C_{1}$. This yields the following expressions for $\sigma (l)$
and $\phi (l)$.

\begin{eqnarray}
\sigma - {\sigma}_{0} = r(l) +{\frac{1}{2}}\ln {\frac{r(l) -
b_{0}}{r(l) + b_{0}}} \qquad  \forall \quad l \neq 0 \nonumber \\
          = 0 \qquad  for \quad l = 0
\end{eqnarray}

where $r(l) = \sqrt{b_{0}^{2} + l^{2}}$.

Also,

\begin{eqnarray}
\phi - {\phi}_{0} = {\frac{1}{2}}\ln {\frac{r(l) -
b_{0}}{r(l) + b_{0}}} \qquad \forall l\neq 0 \nonumber \\
= 0 \qquad for \quad  l = 0
\end{eqnarray}

One can also obtain an expression for $l(\phi)$ for $l\neq 0$.
This is given as :

\begin{equation}
l = \pm b_{0}({\sinh ({\phi}_{0} - \phi)})^{-1}
\end{equation}

In this configuration the value of $\phi$ changes as we move
along the wormhole by varying $l$. At $l=0$ i.e at the throat
$\phi = {\phi}_{0}$, $\sigma = {\sigma}_{0}$. As $l\rightarrow
\pm \infty$ also $\phi \rightarrow {\phi}_{0}$ but $\sigma
\rightarrow \pm \infty$. Thus the string may lie in the wormhole
by spiralling around the surface if one chooses $\phi _{0}$
appropriately.

\subsection{\bf Closed Strings}

A closed, stationary string configuration can be defined in the
following way :

\begin{equation}
t = \tau \quad , \quad l = l_{0} \quad , \quad \theta =
\frac{\pi}{2} \quad , \quad  \phi = C_{1}\sigma
\end{equation}

where $l_{0}$ and $C_{1}$ are constants. A very simple closed string
solution in the background of a wormhole can be defined as

\begin{equation}
t = \tau \quad , \quad l = 0 \quad , \quad \theta =
\frac{\pi}{2} \quad , \quad \phi = C_{1}\sigma
\end{equation}

If $b_{0}$ is the throat radius of the wormhole then one can
easily check that one needs $b_{0} = \frac{1}{C_{1}}$ in order
to satisfy the constraint equation. This, at least in the case
of the Ellis geometry is a possible string configuration.

One recalls from [7] that closed strings are not possible in a
stationary black hole background, such as the Schwarzschild.
The obvious question that arises is --- How does a wormhole
background help in having closed strings ?
In order to understand this we have to put back the $-{\chi}^{2}
(l)$ factor in front of the coefficient of $dt^{2}$ in the
background wormhole metric. This will illustrate the role
of the horizon in forbidding the existence of closed strings.
The string equations of motion for the coordinates $\theta$ and
$\phi$ remain the same as in (15) and (16) while that for the
coordinate  $l$ becomes :

\begin{equation}
l^{\prime \prime} - {\chi} {\chi}^{\prime} - {r\tilde
r}{{\theta}^{\prime}}^{2} - {r\tilde r}\sin^{2}{\theta}{{\phi}^{\prime}}
^{2} = 0
\end{equation}

The constraint equation now changes to :

\begin{equation}
-{\chi}^{2} + {l^{\prime}}^{2} + r^{2}{{\theta}^{\prime}}^{2}
+ r^{2}\sin^{2}{\theta}{{\phi}^{\prime}}^{2} = 0
\end{equation}

Now for any wormhole geometry we have $\tilde r = 0$ at the
throat (i.e. at $l=0$). Therefore, for $\theta =
{\frac{\pi}{2}}$ one gets the following requirements on
$\chi (l)$.

\begin{equation}
\left (\frac{d}{dl} [\chi^{2}(l)]\right )_{l=0} = 0 \qquad ,
\qquad  {\chi}^{2}(l = 0) = b_{0}^{2}C_{1}^{2}
\end{equation}

Note that if $\chi (l = 0) = 0 $ --which is the situation
in the Schwarzschild black hole at the horizon one does not have a
closed string
solution. In general, however, for {\em any } wormhole (which is a
spacetime {\em free of horizons} and singularities) one is bound
to have {\em at least one } closed string configuration residing at the
throat.

\section{PERTURBATIONS AROUND SPECIFIC STRING CONFIGURATIONS}

In this section, we shall concentrate on analysing the nature of
perturbations around two of the specific string configurations
derived in the previous section. A manifestly covariant
formalism for the study of these perturbations has been recently
set up by Larsen and Frolov [8]. The variations are defined as

\begin{equation}
\delta x^{\mu} = \delta x^{i} n^{\mu}_{i} + \delta
x^{a}x^{\mu}_{,a}
\end{equation}

where $n^{\mu}_{i}$ are two vectors normal to the worldsheet
described by the solution. Infact $\{{x}^{\mu}_{,a} ,
{n}^{\mu}_{i}\}$ forms a basis at every point on the worldsheet
($a=\sigma , \tau ; i= 1, 2 $). These vectors obviously satisfy
\begin{equation}
{g}_{\mu \nu}n^{\mu}_{i}n^{\nu}_{j} = {\delta}_{ij}
\qquad , \qquad {g}_{\mu \nu}{x}^{\mu}_{a}{n}^{\nu}_{i} = 0
\end{equation}

We will only consider normal variations because the tangential
ones leave the action invariant -- a consequence of the
reparametrization invariance of the worldsheet. Details and the
exact expressions for the second variation of the Nambu--Goto
action can be found in the paper by Larsen and Frolov [8]. Since
we consider only stationary strings our equations will br much
simpler  than those for the most general case. It turns out that
the perturbations are governed by a set of equations written in
compact form as :

\begin{equation}
\left ( {{\partial}_{{\sigma}_{c}}}^{2} - {{\partial}_{\tau}}^{2} \right
)  {\delta} x_{i} = U_{ij}{\delta}x^{j}
\end{equation}

where $U_{ij}$ is given by

\begin{equation}
U_{ij} = V{\delta}_{ij} + {F}^{2}V_{ij}
\end{equation}

and
\begin{equation}
V = {\frac{1}{4}}\left ({F}^{\prime 2} - 2F{F^{\prime \prime}}\right )
\end{equation}

\begin{equation}
V_{ij} = x^{p\prime}x^{s\prime}{\tilde
R}_{pqrs}{{n}^{q}_{i}}{{n}^{r}_{j}}
\end{equation}

with $d{\sigma}_{c} = \frac{d\sigma}{F}$, $F \equiv {\chi}^{2}$,
and $(n^{p}_{i},x^{\prime i})$ form an orthonormal basis in the
space with metric $H_{ij}$.

Thus, to obtain information regarding the nature of perturbations
one needs to solve the coupled differential equations involving
the ${\delta}x^{i}$. We now concentrate on doing so for the
specific string configurations discussed in Section II.

\subsection{\bf Straight String}

The normal vectors for this case can be chosen as :

\begin{equation}
{n}^{\mu}_{1} \equiv (0, 0, {\frac{1}{r}}, 0) \qquad ; \qquad
n^{\mu}_{2} \equiv (0, 0, 0, {\frac{1}{r}})
\end{equation}

and the perturbations in terms of ${\delta}x_{1},{\delta}x_{2}$
are

\begin{equation}
\delta l = 0 \quad, \quad \delta \theta = {\delta}x_{1}/r \quad ,
\quad {\delta}\phi = {\delta}x_{2}/r
\end{equation}

Using the expression for $V_{ij}$ we find that:

\begin{equation}
V_{11} = V_{22} = {\frac{b_{0}^{2}}{(b_{0}^{2} + l^{2})^{2}}} \qquad ;
\qquad V_{12}=V_{21} = 0
\end{equation}

Since $F = {\chi}^{2} = 1$ we have $V = 0$.

Hence we need to solve only one differential equation in order
to obtain ${\delta}x_{1}$ and ${\delta}x_{2}$. This is given as

\begin{equation}
\left ( {\partial}^{2}_{\sigma} - {\partial}^{2}_{\tau} \right ) \delta
x_{1,2} = {\frac{b_{0}^{2}}{(b_{0}^{2} + l^{2})^{2}}} \delta x_{1,2}
\end{equation}

Now we perform  an usual Fourier expansion of
${\delta}{x_{1,2}}$. This yields the equation

\begin{equation}
\frac{d^{2}D_{\omega}}{{d\sigma}^{2}} + \left ( {\omega}^{2} -
 {\frac{b_{0}^{2}}{(b_{0}^{2} + l^{2})^{2}}} \right ) D_{\omega}
=0
\end{equation}

where

\begin{equation}
{\delta}x_{1,2} = \int e^{-i\omega t} D_{\omega}(\sigma)d\omega
\end{equation}

This differential equation for $D_{\omega}$ can be exactly
solved. Infact it appears in the study of massless scalar waves
in the background of the Ellis geometry  and has been analysed
in detail in [16]. Redefining $D_{\omega} = \sqrt{b_{0}^{2} +
{\sigma}^{2}} F_{\omega}$ and with $\xi = \frac{\sigma}{b_{0}}$
we get

\begin{equation}
(1+{\xi}^{2}) {\frac{d^{2}F_{\omega}}{d{\xi}^{2}}} + 2\xi
\frac{dF_{\omega}}{d\xi} + [ {\omega}^{2}
b_{0}^{2}(1+{\xi}^{2})]F_{\omega} ] = 0
\end{equation}

The differential equation for Radial Oblate Spheroidal Functions is

\begin{equation}
(1+{\xi}^{2}) {\frac{d^{2}V_{mn}}{d{\xi}^{2}}} + 2\xi
\frac{dV_{mn}}{d\xi} + [ - {\lambda}_{mn} + k^{2}{\xi}^{2}
- {\frac{m^{2}}{1+{\xi}^{2}}} ] V_{mn} = 0
\end{equation}

If we take $m=0$ ${\lambda}_{0n} = -k^{2} =
-{\omega}^{2}b_{0}^{2}$ then we get our equation governing the
perturbations of a straight string as a special case of the
Radial Oblate Spheroidal Equation. However we only get solutions
for specific values of ${\lambda}_{0n}$ (i.e ${\lambda}_{0n}$
negative). Consulting [17] we notice that this is possible only
for $n=0,1$. The solutions to the equations are finite at
infinity -- they behave like simple sine/cosine waves. The scattering
problem for the Schroedinger--like equation has been analysed
numerically in [18] and one obtains a smooth transmittivity curve
rising from almost zero at small values of ${\omega}$ and
increasing to one as the energy becomes comparable or larger
than  the barrier height. Explicit expressions for the solutions
of the equations governing the perturbations can be written only
for $n=0,1$ and these involve a series of products of Bessel
Functions (for details see [17]).

\subsection{\bf Closed String}

The equations governing the perturbations of the closed string
are even more simple. The normals can be chosen as :

\begin{equation}
{n}^{\mu}_{1} \equiv (0, 0, {\frac{1}{r}}, 0) \qquad ; \qquad
n^{\mu}_{2} \equiv (0, 1, 0, 0)
\end{equation}

and the ${\delta}x^{\mu}$ turn out to be

\begin{equation}
\delta l = \delta x^{2} \quad , \quad  \delta\theta = {\delta
x^{1}}/r \quad , \quad  \delta\phi = 0
\end{equation}
The potentials are either constant or zero (by virtue of the
fact that $l = 0$).

\begin{equation}
V_{11} = - V_{22} = C_{1}^{2} \qquad ; \qquad V_{12} = V_{21} =0
\end{equation}

Performing a Fourier expansion similar to the case discussed
earlier one arrives at the following expressions for
$D^{1}_{\omega}$ and $D^{2}_{\omega}$. These are:

\begin{equation}
D^{1}_{\omega}(\sigma) = A_{1}\exp(\pm in\sigma)
\end{equation}

where ${\omega}_{n}^{2} =  n^{2} - C_{1}^{2}$ and

\begin{equation}
D^{2}_{\omega}(\sigma) = A_{2}\exp(\pm in\sigma)
\end{equation}

where  ${\omega}_{n}^{2} =  n^{2} + C_{1}^{2}$

The $\tau$ parts of the $\delta x^{1,2}$ differ in the argument
of the exponential. For $\delta x^{1}$ it is $\exp(\pm
i\sqrt{n^{2}-C_{1}^{2}}\tau)$ whereas for $\delta x^{2}$ it is $\exp(\pm
i\sqrt{n^{2}+C_{1}^{2}}\tau)$. Notice that for $\delta x^{1}$
the $\tau$ part may be exponentially damped or growing if
$C_{1}^{2} \ge n^{2}$ whereas the solution for $\delta x_{2}$
is always oscillatory.

\section{ THE GENERALIZED RAYCHAUDHURI
EQUATIONS}

The Raychaudhuri Equations in GR [19] essentially deal with the
evolution of timelike or null geodesic congruences. The
assumption of an Energy Condition  leads to the focussing
theorem in GR which states that if matter satisfies certain
restrictions then initially converging sets of geodesics will
always tend to focus at a point within a finite value of the
affine parameter that characterises each geodesic. Recently
Capovilla and Guven [9] have come up with a generalisation of the
Raychaudhuri Equations for $D$ dimensional timelike  worldsheets
embedded in an $N$ dimensional curved background. We shall be
concerned with the $D=2,N=4$ case of these generalised
Raychaudhuri Equations.

We now very briefly recall the ingredients of the generalised
Raychaudhuri equations (for details the reader is referred to
Capovilla and Guven[9]).The equation which we shall deal with
is given by (this is a special case of the more general equation
quoted in [9]):

\begin{equation}
\Delta \gamma + {\frac{1}{2}}{\partial}_{a}\gamma{\partial}^{a}\gamma +
(M^{2})^{i}_{i} = 0
\end{equation}

where ${\nabla}_{a}$ is the worldsheet covariant derivative
, and ${\partial}_{a}\gamma={\theta}_{a}$. ${\theta}_{a}$ is
related to a quantity $J_{a}^{ij} = {\frac{1}{2}}
{\delta}_{ij}{\theta}_{a}$ where the $J_{a}^{ij}$ are defined
as quantittes related to the normal gradients of the orthonormal
spacetime basis $(E_{a}^{\mu},n^{\mu}_{i})$ defined at each point
on the worldsheet through the analogs of the Gauss--Weingarten
equations :

\begin{eqnarray}
D_{i}E_{a} = J_{aij}n^{j} + S_{abi}E^{b} \\
D_{i}n^{j} = -J_{aij}E^{a} + {\gamma}^{k}_{ij}n^{k}
\end{eqnarray}

$S_{abi}$ and ${\gamma}^{k}_{ij}$ are defined as :

\begin{eqnarray}
S^{i}_{ab} = {g}_{\mu\nu}n^{\alpha
i}(D_{\alpha}E^{mu}_{a})E^{\nu}_{b} \\
{\gamma}^{k}_{ij} = {g}_{\mu\nu}n^{\alpha
i}(D_{\alpha}n^{\mu}_{j})n^{\nu}_{k}
\end{eqnarray}

The quantity $(M^{2})^{ij}$ is given as (see [20]):

\begin{equation}
(M^{2})^{ij} = K_{ab}^{i}K^{abj} +
R_{\mu\nu\rho\sigma}E^{\mu}_{a}n^{\nu i}E^{\rho a}n^{\sigma j}
\end{equation}

Notice that the quantity $(M^{2})^{ij}$ contains contributions
from the extrinsic curvatures $K_{ab}^{i} =
-g_{\mu\nu}(D_{a}E^{\mu}_{b}) n^{\nu}$ of the worldsheet as well as the
Riemann tensor components of the background spacetime. For
geodesic curves one can arrive at the usual Raychaudhuri
equation by noting that $K^{i}_{00} = 0$, the $J_{aij}$ are
related to their spacetime counterparts $J_{\mu\nu a}$ through
the equation $J_{\mu\nu a} = n^{i}_{\mu}n^{j}_{\nu}J_{aij}$,
and the $\theta$ is defined by contracting with the projection
tensor $h_{\mu\nu}$.

The $\theta _{a}$ or $\gamma$ basically tell us how the
spacetime basis vectors change along the normal directions  as we
move along the worldsheet. If ${\theta}_{a}$ diverges somewhere
, it induces a divergence in $J_{aij}$ , which, in turn means
that the gradients of the spacetime basis along the normals have
a discontinuity. Thus the family of worldsheets meet along a
curve and a cusp/kink is formed. This can be called as a
worldsheet focussing effect in analogy with geodesic focussing in GR.

We emphasize that the
simplified form of the generalised Raychaudhuri equation we are using
is valid only when the quantity $(M^{2})^{ij}$ does not contain
any off diagonal terms. Quantities analogous to the shear and
rotation present in the case of geodesic congruences are also
present here but we have put them to zero in order to simplify
the analysis.

We now move on to the discussion of the special cases.

\subsection{\bf Straight String}

The induced metric $\gamma_{ab} =
{g}_{\mu\nu}x^{\mu}_{,a}x^{\nu}_{,b}$ is equal to ${\eta}_{ab}$
here-- thus the $K_{ab}^{i}$ vanish for all $a,b$. The normal
vectors can be chosen as before. The quantity $(M^{2})^{i}_{i}$
turns out to be

\begin{equation}
(M^{2})^{i}_{i} = -\frac{2b_{0}^{2}}{(b_{0}^{2} + {\sigma}^{2})^{2}}
\end{equation}

Thus the Raychaudhuri equation for $\gamma = 2\ln\beta$ turns
out to be:

\begin{equation}
-\frac{{\partial}^{2}\beta}{{\partial}{\tau}^{2}} +
\frac{{\partial}^{2}\beta}{{\partial}{\sigma}^{2}}
- \frac{2b_{0}^{2}}{(b_{0}^{2}+{\sigma}^{2})^{2}}\beta =0
\end{equation}

Separating variables $\beta = T(\tau)\Sigma (\sigma)$ we have

\begin{equation}
\frac{d^{2}T}{d{\tau}^{2}} + {\omega}^{2} T = 0
\end{equation}

and

\begin{equation}
\frac{d^{2}\Sigma}{d{\sigma}^{2}} + \left ( {\omega}^{2} - \frac
{2b_{0}^{2}}{(b_{0}^{2} + {\sigma}^{2})^{2}} \right ) \Sigma = 0
\end{equation}

Since $ \gamma = 2\ln\beta = 2\ln T + 2\ln {\Sigma}$
we have:

\begin{equation}
{\theta}_{\tau} = 2\frac{\dot T}{T} \qquad ; \qquad
{\theta}_{\sigma} = 2\frac{{\Sigma}^{\prime}}{\Sigma}
\end{equation}

Now, from several theorems in the theory of differential equations
[21] one can conclude that as long the quantity $H(x)$ in
the differential equation $\frac{d^{2}F}{dx^{2}} + H(x)F = 0$
is positive and continuous everywhere in the domain of $x$ one
can say that there exists zeros in the solutions. If we use this
fact as an input in the two differential equations for $T$ and
$\Sigma$ we find that both $\theta_{\tau}$ and $\theta_{\sigma}$
must necessarily diverge somewhere. For $T(\tau)$ we can locate
the points wheresas for $\Sigma$ the only statement we can make
is ${\omega}^{2} \ge \frac{2}{b_{0}^{2}}$ (this guarantees the
positivity of $H(\sigma)$ )

\subsection{\bf Closed String}

Similarly for the case of closed strings the generalised Raychaudhuri
Equations can be written down.
We find

\begin{equation}
(M^{2})^{i}_{i} = 0
\end{equation}

and

\begin{equation}
-\frac{{\partial}^{2}\beta}{{\partial \tau}^{2}}
+\frac{{\partial}^{2}\beta}{{\partial \sigma}^{2}} = 0
\end{equation}

Using analysis similar to the previous case one obtains the
following solutions for $T(\tau)$ and $\Sigma (\sigma)$:

\begin{equation}
T(\tau) = \sin n \tau
 \qquad ;
\qquad  \cos n\tau
\end{equation}

\begin{equation}
\Sigma (\sigma) = \sin n\sigma \qquad ; \cos n\sigma
\end{equation}

The conclusions on focussing are similar  to the case for
Straight strings.

\subsection{Strings with $\phi \neq constant$}

Finally we write down the corresponding Raychaudhuri equations
for the string solutions with a nonconstant $\phi$. These, after
the usual separation of variables turn out to be the following
two ordinary differential equations for $\Sigma$ and $T$.

\begin{equation}
\frac{d^{2}\Sigma}{d{\sigma}^{2}} + \left ( {\omega}^{2} -
\frac{2l^{2}(\sigma)b_{0}^{2}}{r^{6}(\sigma)} \right ) \Sigma = 0
\end{equation}

\begin{equation}
\frac{d^{2}T}{d{\tau}^{2}} + {\omega}^{2}T = 0
\end{equation}

The second equation is the same as for the previous two cases.
For the first, one has to check the positivity of the quantity $H(\sigma)$
. This can be done by locating the maximum value of the
potential function which occurs here at $l = \frac{b_{0}}{2\sqrt
2}$. Thus ${\omega}^{2} \ge \frac{128}{729b_{0}^{2}}$ is the
criterion for the positivity of $H(\sigma)$. Thus a focussing
effect is possible in the sense that there exist zeros in the
solutions of the differntial equations and $\theta_{a}$ diverges
at certain points.

\section{CONCLUDING REMARKS}

Let us now summarize the results we have obtained.

(i) \quad Specific string configurations in a wormhole
background have been obtained. These include both closed and
open strings. The open ones can be straight or spiral while the
closed one is possible only at the throat of the wormhole. All
geodesics on the two--sphere residing at the throat of the wormhole
are closed string configurations. In fact it should be
emphasized that for all wormholes there exists at least one
closed string configuration at the throat. It is important to contrast this
with the case of the black hole where no such closed string
configurations are possible. This fact can be attributed to the
presence of the black hole horizon. We have also discussed a
general solution  given in terms of elliptic functions

(ii) \quad Perturbations about two of the string configurations
have been analysed using the manifestly covariant formalism
of Larsen and Frolov [8]. For the straight string the
perturbation equations can be solved exactly in terms of the
Radial Oblate Spheroidal Functions. For closed strings the
solutions are even simpler --they are only sines and cosines.

(iii) \quad The Raychaudhuri Equations for deformations of the
string worldsheets obtained in Section I have also been written
down and analysed for the various configurations.
We have been able to arrive at the concept of worldsheet
focussing in a manner similar to geodesic focussing in GR.
In the general case of a string moving in any curved background
one can see very easily that the generalised Raychaudhuri equation (with
quantities analogous to shear and rotation put to zero)
takes the following form:

\begin{equation}
- \frac{{\partial}^{2}F}{\partial{\tau}^{2}} +
\frac{{\partial}^{2}F}{\partial{\sigma}^{2}}
+\Omega^{2}(\sigma,\tau)  (M^{2})^{i}_{i}(\sigma,\tau)F = 0
\end{equation}

This is possible because the worldsheet is two--dimensional
and any $2D$ metric can be written in a conformally flat
form by choosing coordinates appropriately. The operator
$\Delta$ therefore reduces to $\frac{1}{{\Omega}^{2}}\Delta_{M}$
where $\Delta_{M}$ is the D'Alembertian in flat Minkowski space.
and $\Omega^{2}$ is the conformal factor. Therefore, to derive a
focussing theorem for string worldsheets
one has to understand the nature of the solutions of the above
equation with emphasis on the zeros.

(iv)\quad Since a closed string exists only at the
throat and the wormhole requires matter that violates the Weak
Energy Condition one can ask -- Can quantum strings provide
the source for WEC violating matter? If this is true then one
has a solution to the problem of WEC violation for wormhole
material.

Finally, as a continuation of the search for exact solutions of
the string equations and constraints in curved spacetimes one
can look into string motion in other wormhole backgrounds.
An interesting case could be the evolving wormholes
,where the analysis used here would not apply
but that of string configurations in
Friedman--Robertson--Walker cosmologies certainly would [21].

Work along these lines is in progress and will be communicated
in due course.

\section{ACKNOWLEDGEMENTS}

It is a pleasure to thank Alok Kumar
and Jnanadeva Maharana for useful discussions.
Financial support from the Institute of
Physics, Bhubaneswar, in terms of a Fellowship
is also gratefully acknowledged.

\pagebreak

\centerline{\bf REFERENCES}

\begin{itemize}
\begin{enumerate}
\item H. J. de Vega and N. Sanchez, Phys. Rev. {\bf D42}, 3969 (1990)
; H. J. de Vega, M. Ramon--Medrano and N. Sanchez, Nucl. Phys.
{\bf B374}, 405 (1992)
\item H. J. de Vega and N. Sanchez, Phys. Rev. {\bf D47}, 3394
(1993)
\item V. P. Frolov, V. D. Skarzhinsky, A. I. Zelnikov and O. Heinrich
, Phys. Letts. {\bf B224} (1989)
\item H . J. de Vega, A. L. Larsen and N. Sanchez, Nucl. Phys.
{\bf B427}, 643 (1994); H . J. de Vega, A. L. Larsen and N. Sanchez,
Preprint--DEMIRM-- Obs de Paris--94049
\item H. J. de Vega and N. Sanchez, Nucl. Phys. {\bf B317}, 706 (1989)
{\em ibid} 731 (1989)
; Phys. Letts. {\bf B244}, 215 (1990); Phys. Rev. Letts. {\bf
65C}, 1517 (1990); Int. Jour. Mod. Phys. {\bf A7}, 3043 (1992);

D. Amati and C. Klimcik, Phys. Letts. {\bf B210}, 92 (1988) ;

M. Costa and H. J. de Vega, Ann. Phys. {\bf 211}, 223 (1991)
{\em ibid} 235 (1991)

C. Lousto and N. Sanchez, Phys. Rev {\bf D46}, 4520 (1992)
\item H. J. de Vega, A. V. Mikhailov and N. Sanchez, Teor. Mat.
Fiz {\bf 94}, 232 (1993); F. Combes, H. J. de Vega, A.V
Mikhailov and N. Sanchez, Phys. Rev {\bf D50}, 2754 (1994)
\item A. L. Larsen and N. Sanchez, Preprint--hep--th 9501101
\item A. L. Larsen and V. P. Frolov, Nucl. Phys {\bf B414}, 129
(1994)
\item R. Capovilla and J. Guven, Preprint CIEA--GR--9409, ICN--UNAM--9406
(gr--qc 9411061)
\item M. S. Morris and K. S. Thorne, Am. Jour. Phys {\bf 56},
395 (1988)
\item M. S. Morris, K. S. Thorne and U. Yurtsever, Phys. Rev.
Letts. {\bf 61}, 1466 (1988)
\item J. L. Friedman, M. S. Morris, I. D. Novikov, F. Echeverria
, G. Klinkhammmer, K. S Thorne and U. Yurtsever, Phys. Rev. {\bf
D 42}, 1915 (1990); S. W. Kim and K. S. Thorne, Phys. Rev {\bf D
43}, 3929 (1991)
\item C. W. Misner, K. S. Thorne and J. A. Wheeler, Gravitation
(W. H Freeman, San Francisco, 1973)
\item H. Ellis, J. Math. Phys. {\bf 14}, 104 (1973)
\item R. M. Wald, General Relativity (University of Chicago Press
,1985)
\item S. Kar, D. Sahdev and B. Bhawal, Phys. Rev {\bf D49} 853
(1994)
\item {\em Handbook of Mathematical Functions with Formulas,
Graphs and Mathematical Tables}, edited by M. Abramowitz and I. A. Stegun
(Dover, New York, 1965)
\item S. Kar, S. N. Minwalla, D. Mishra and D. Sahdev (to appear
in Phys. Rev {\bf D50})
\item A. K. Raychaudhuri, Phys. Rev {\bf 98}, 1123 (1955)
\item {\em Th expression for $(M^{2})^{ij}$ given above does not match
with the corresponding one in [9]--a sign error in the Ricci Identity
used there being the root cause behind this. This fact can also be noticed
by comparing the perturbative Jacobi equations in the penultimate section
in [9] with the corresponding one in the paper by Larsen and Frolov [8]}
\item F. J. Tipler, Ann. Phys {\bf 108},1 (1977)
\item A. L. Larsen and N. Sanchez, Preprint--hep--th 9501102
\end{enumerate}
\end{itemize}

\pagebreak

\appendix
\section*{ Affine Connections and Riemann Tensors}

In this appendix we list all the relevant affine connections and
Riemann tensors used in the calculations in this paper.

The metric is assumed as:

\begin{equation}
ds^{2} = -\chi^{2}(l)dt^{2} + dl^{2} + r^{2}(l)\left (d{\theta}^{2}
+\sin^{2}\theta d\phi^{2} \right )
\end{equation}

the nonzero Affine Connections (Christoffel symbols) are given as

\begin{equation}
{\Gamma}^{t}_{lt} = \frac{{\chi}^{\prime}}{\chi}\quad ; \quad
{\Gamma}^{l}_{tt} = \chi {\chi}^{\prime} \quad ; \quad
{\Gamma}^{l}_{\theta\theta} = -rr^{\prime}
\end{equation}

\begin{equation}
{\Gamma}^{l}_{\phi\phi} = -rr^{\prime}\sin^{2}\theta
\quad;\quad {\Gamma}^{\theta}_{\theta l} = \frac{r^{\prime}}{r}
\quad ;\quad {\Gamma}^{\theta}_{\phi\phi} =
-\sin{\theta}\cos{\theta}
\end{equation}

\begin{equation}
{\Gamma}^{\phi}_{l\phi} = \frac{r^{\prime}}{r}\quad ;\quad
{\Gamma}^{\phi}_{\phi\theta} = \cot \theta
\end{equation}

{}From these one can evaluate the Riemann tensor components which
are:

\begin{equation}
R^{l}_{\theta l \theta} = R^{l}_{\phi l \phi} =
-rr^{\prime \prime} \quad ;\quad R^{\theta}_{\phi \theta \phi} =
\sin^{2}\theta (1-{r^{\prime}}^{2})
\end{equation}

In the above expressions the prime denotes derivative w.r.t $l$.
With the help of these one can easily evaluate the relevant
expressions for Ellis geometry ($\chi = 1$ and $r(l) =\sqrt
{b_{0}^{2} + l^{2}}$). These are

\begin{equation}
{\Gamma}^{l}_{\theta\theta} = {\Gamma}^{l}_{\phi \phi} = -l
\quad ; \quad {\Gamma}^{\theta}_{\theta l} = {\Gamma}^{\phi}_{l
\phi}  = \frac{l}{b_{0}^{2} + l^{2}}
\end{equation}

${\Gamma}^{\theta}_{\phi \phi}$ and ${\Gamma}^{\phi}_{\phi
\theta}$ are the same as quoted above.

\begin{equation}
R^{l}_{\theta l \theta} = R^{l}_{\phi l \phi} =
-\frac{b_{0}^{2}}{b_{0}^{2} + l^{2}} \quad ; \quad
R^{\theta}_{\phi \theta \phi} =
\frac{b_{0}^{2}\sin^{2}\theta}{b_{0}^{2} +l^{2}}
\end{equation}

\end{document}